\documentclass[twocolumn,twoside,slac]{revtex4}
\usepackage{graphicx}
\usepackage{fancyhdr}
\begin{document}

\title{The Redesigned BaBar Event Store: Believe the Hype}

%

\author{Adeyemi Adesanya}
\email{yemi@slac.stanford.edu}
\author{Jacek Becla}
\email{becla@slac.stanford.edu}
\author{Daniel Wang}
\email{danielw@slac.stanford.edu}
\affiliation{SLAC, Stanford, CA 94025, USA}

\author{\vspace{2ex} On behalf of the BaBar Computing Group}
\noaffiliation

\begin{abstract}

As the BaBar experiment progresses, it produces new and unforeseen
requirements and increasing demands on capacity and feature base.
The current system is being utilized well beyond its original design
specifications, and has scaled appropriately, maintaining data consistency
and durability.  The persistent event storage system has remained largely
unchanged since the initial implementation, and thus includes many design
features which have become performance bottlenecks.  Programming
interfaces were designed before sufficient usage information became
available.  Performance and efficiency were traded off for added
flexibility to cope with future demands.  With significant
experience in managing actual production data under our belt, we are now
in a position to recraft the system to better suit current needs.  The
Event Store redesign is intended to eliminate redundant features while
adding new ones, increase overall performance, and contain the physical
storage cost of the world's largest database.

\end{abstract}

\maketitle

\thispagestyle{fancy}


\section{Introduction} \label{sec-intro}
The purpose of the event store is to provide durable persistence for
physics event data~\cite{evsdesign-ref}~\cite{jacek-old-ref}.  The system
must scale as the experiment continues, 
which currently means billions of physics events organized in millions of
collections.  All data must be persisted, and none of it may be thrown
away.  The system must allow access to data generated at any point in the
experiment-- from its inception to what's currently being generated.  

The system should always be available.  It must provide reliable and
robust operation without regular outages.  Since the system is available to
collaborators from over 70 institutions around the world, any outage is
disruptive.  

Data analysis and other high-level access is separated from event store
persistence via an abstraction layer.  This split in dependency permits
the asynchronous development in transient and persistent code that is
necessary in such a large and complex system.  Without such a split, this
redesign could be far more disruptive.

This paper describes the redesign of the BaBar Event Store, detailing
the current situation, the motivation, the techniques, and implementation.
Section~\ref{sec-intro} provides an overview of the paper, and
section~\ref{sec-past} describes the current
system. Section~\ref{sec-general} details the motivation and overall
design of the redesign project.  Section~\ref{sec-impl} details the
implementation, and Sections \ref{sec-impt} and \ref{sec-status} describe
the estimated impact of this redesign and the status of this project,
respectively.  Section~\ref{sec-concl} summarizes the project.
\section{Past Work} \label{sec-past}

\subsection{ODBMS back-end}
The current, production BaBar event store is written on top of an
Objectivity/DB object-oriented database management system
(ODBMS)~\cite{objy-ref}.
Using the Objectivity persistence gives our system useful primitives for
atomic transactions, consistent fault-tolerant state, and durable storage
that is designed to survive software and hardware faults.  

\subsection{Abstracted Persistence}
The current system abstracts the persistence layer in an effort to reduce
dependence on the particulars of the underlying database implementation.
This strategy stabilizes the client code against upgrades and other
version changes of the persistent system.

\subsection{Flexible Architecture}
The current system was designed to accomodate the demands of a new
experiment that had not established  specific needs or usage
characteristics.  The designers planned for these unknowns by building
flexibility in many parts of the system, thus allowing the system to
evolve as requirements and demands evolved.  For event storage, this
flexibility is centered around the concept of an ``event.''

\subsubsection{Transient Events}
Transient events are represented very generally as typed bags of objects.
The transient structure enforces nothing but this.  Arbitrary types of
data objects can be inserted in transient events with or without
keys to identify them.  This flexible interface is what is exported to
high-level analysis code.
In practice, objects stored in transient events
can be categorized as: identification data, analysis data (``micro'' and
``mini'' levels), and event store metadata.  Identification data provides
information about the event, i.e. the when? and where? of the event.
Analysis data includes actual event data to be analyzed, i.e. the what?
of the data.  Event store metadata includes objects stored in the
transient event to aid in conversion to and from persistence.  

\subsubsection{Persistent Events}
The flexibility allowed by the transient event structure implies a certain
level of flexibility in the persistent layer.  An important vehicle of
this flexibility is the concept of event headers, which provide
indirection between the navigational and data parts of events, insulating
both sides from each other's code development.  In the current system, the
headers are actual persistent structures.

\subsubsection{Event Management}
To manage and make sense of billions of physics events, the event store
exports an organizational concept of {\it collections} of events.
Collections are named sets of events.  A particular event is part of one
or more collections, although it is only ``owned'' by one.
Collections are identified through a hierarchical naming system, which
allows the system to export some lightweight access control.

\section{The Redesign} \label{sec-general}
\subsection{Motivation}
With a successful production system in place, what are the motivations for
redesigning the system?  The answers are threefold: \emph{cost}, 
\emph{performance}, \emph{features}. 

\subsubsection{Cost} 
The event store currently grows at an estimated rate of 500GB per
day~\cite{jacek-current-ref}.  As
the experiment continues, upgrades are made in data collecting, and the
number of events collected grows.  If past experience is any indication,
this number grows faster than the famous Moore's
law.  The redesign aims to stem the resulting deluge of data by reducing
the amortized footprint per-event.   This is the primary reason for the
redesign.  

\subsubsection{Performance}
The production system includes great flexibility, based on assumptions and
expectations based on previous HEP experiments.  Some aspects of this hailed
flexibility had implementations that were costly in terms of both space
and performance.  By streamlining the persistent event structure and
removing unneeded indirection, the redesign should yield significant gains
in performance. 

\subsubsection{Features} 
With the needs of maintaining a stable production system a top priority,
some features cannot be added without significant code changes.  The
redesign provides a convenient point at which a few important features may
be added.\footnote{Details of these new features are specific and beyond
the scope of this paper. An example is the introduction of a common DB ID
object, which should aid system administration tremendously.}
\subsection{Design}
The redesign aims to utilize the accumulated administrative and
maintenance experience of the running system to produce an optimized
persistency system.  Simple changes that produce simpler and more
maintainable code are
preferred over more elaborate designs.  To be clear, the BaBar database
group is not a physics analysis group by any definition, so the redesign is
tightly focused on the structural parts of the event that are hidden to
analysis. 

\subsubsection{Share redundant data} 
A lot of data is identical in successive events.  Hand analysis of
production events has identified a subset of fields in event objects that are
changing slowly, if at all.\footnote{``Slowly changing'' here means
changing approximately every thousand events.}  Sharing these fields
should result in substantial space savings.
\subsubsection{Eliminate unused data}
Some event fields are obsolete.  Though we are not the experts on event
data, we 
have identified a few parts to be obsolete with the help of physicists in
BaBar.  Some fields and data structures were borrowed from previous
experiments and thus are good candidates for elimination.
\subsubsection{Reorganize data into more efficient structures}
Some existing data structures can be restructured into significantly more
efficient structures, in terms of size and performance.  For example,
data structures whose contents rarely change can be stored more tightly,
saving on precious persistent footprint, at the cost of more intelligent
access code.  In some cases, flexibility will be increased when the data
is structured more appropriately for the usage that has been observed.

\section{Implementation} \label{sec-impl}
\subsection{Overview}
\begin{figure*}[htbp]
\includegraphics[width=135mm]{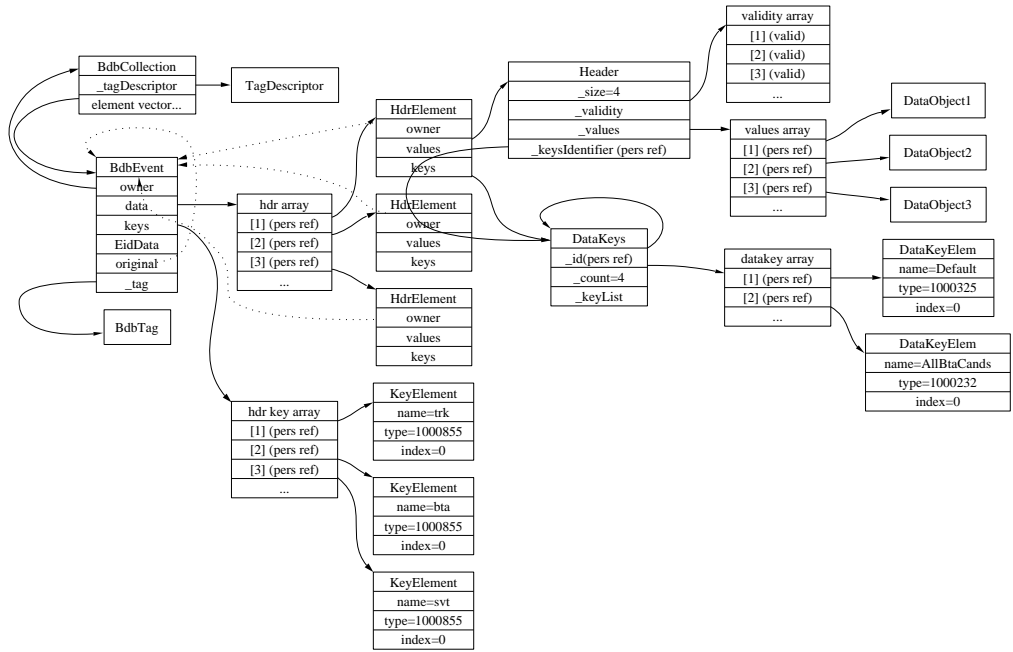} 
\caption{Current persistent event structure\label{fig:bdbevent}}
\end{figure*}

{\bf Figure~\ref{fig:bdbevent}} illustrates the current situation.  The
flexibility of the 
original design is obvious.  Indirection is common, allowing the structure
to scale to large numbers of data objects and arbitrary types without
problems.  The only problem is that its flexibility is heavyweight.  Much
of the structure is redundant, and identical over successive events.
	
\begin{figure*}[htbp]
\includegraphics[width=135mm]{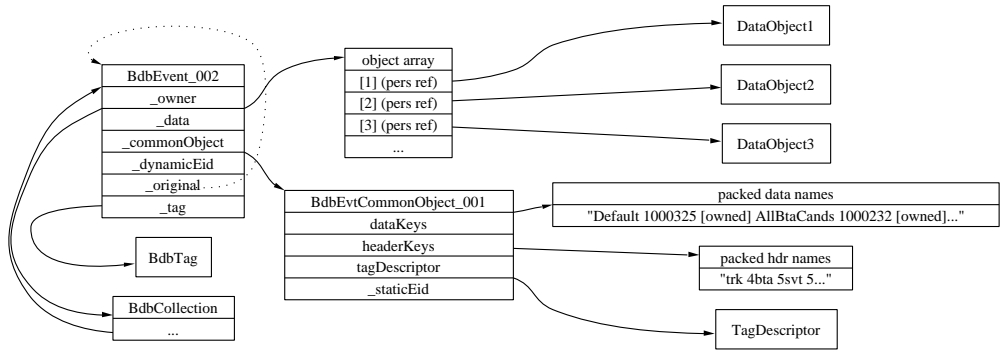}
\caption{Redesigned persistent event structure\label{fig:bdbevent_002}}
\end{figure*}

{\bf Figure~\ref{fig:bdbevent_002} } illustrates the redesigned event model.
Much indirection has been eliminated.  The flexibility of the old system
exists, but has been shared among events, making heavy demands on
flexibility expensive, while the occasional utilization of 
flexibility is small to almost-negligible.

\subsubsection{Event Structure}
The original design allowed essentially the same flexibility in
persistency that the transient system provided.  The redesign takes
advantage of the fact that events and their associated data, once created,
are very rarely changed.  Each event's set of data objects is different in
content, but the \emph{types} of data objects attached to an event are
stable and almost identical over the life of a particular job.
With this in mind, we have altered the persistent event structure to
reuse data where possible, and store data more space efficiently.

\subsubsection{Event Tags}
No single analysis task involves processing the entire contents of the Event
Store. Some form of coarse pre-selection is necessary in order to
efficiently arrange events and collections into logical hierarchies, and
provide a suitable jump-off point for end-user analysis jobs that focus on
a sample of events. Event tags facilitate this coarse pre-selection.  When
an event is reconstructed from raw detector 
information or simulation data, its corresponding event tag is also
created and associated with it.  

{\bf Definition} A Tag
contains arrays of attributes that effectively summarize an event's state.
Tags are intended to be relatively small compared to the overall event size
to facilitate rapid filtering.  Tag attributes (also known as \emph{bits} or
\emph{fields}) have a value of a distinct data type and are identified by a
string
name. 

{\bf Descriptors}
Each Tag is considered to be unique but attribute names are common among
large runs of events so we choose to store these names in a single object
shared among many events. 
This object is called a Tag Descriptor and is
simply a look-up table for Tag attributes. Each attribute name has a unique
key that is used to dereference the arrays in each Tag and access the value
associated with the attribute. 

In the current Event Store, a single Tag
Descriptor is held in each event collection. This seemed like a reasonable
solution back in 1997 but it has some serious implications. Analysis jobs
can read events from multiple input sources and produce new output
collections. These output collections are described as sparse because they
can refer to a diverse group of events that may have very little in common.
Events' Tags may even have different attributes but because they are
stored in the same collection they must share one Tag Descriptor instance.
This results in the creation of bloated Tag objects that contain every single
possible attribute in order to ensure consistency throughout the collection.
Not only does this waste space, it is very misleading since we have no real
way of determining if an attribute is valid for a given Tag.

\subsection{Sharing Redundant Data}
\subsubsection{Common Objects}
Careful analysis of currently persisted events in the system show that
many fields in the event change slowly-- i.e. their values are identical
over hundreds(or more) of events.  Why not place these fields in a single
object, and store a reference to that object in each event with those
fields? 
This is the idea behind the common object.  The current redesign system
allows for a single common object for event data, eschewing different
common objects for different fields (providing a higher granularity of
sharing) because potential gain is limited, and probably smaller than the
extra pointers needed to store, not to mention the added maintenance
complexity. 

Transient events each store a standard object called AbsEventID, which
contains 
fields used to identify an event.  Since BaBar event jobs require, and
thus assume existence of this object, its contents can be stored directly
in the event without incurring the overhead of a separate persistent
object. 
Some fields of this object(the event ID) change often, and others rarely.
The latter fields are placed in the common object.  These sets of fields
are labeled the \emph{dynamic} and \emph{static} fragments of the event
ID. 

\subsubsection{Event Headers}
The multi-leveled ``event header'' structure is flexible and allows simple
updating.  However, the structure involves many persistent objects.  Since
the structure is constant over many objects, it makes sense to store a
compacted representation of the structure instead of the structure
itself.  This makes updating the structure expensive, but since structural
updates are rare, the benefits in size are worth it.  

An event's data objects are stored in ``headers'', which have arbitrary
keys(names in character-string format).  The data objects themselves are
keyed according to key(character-string format again) and type.  In the
redesign, all of an event's data objects are stored in a single array, and
the keying information is stored in packed strings and offloaded to the
common object, where they can be shared.

\subsubsection{Tag Descriptor}
For the redesign, we decided to apply the common object concept by removing
the single Tag Descriptor from the collection and sharing it directly among
Event Tags that have identical attribute lists. There would be no
possibility of adding attributes to a Tag Descriptor. If a matching Tag
Descriptor could not be found for an Event Tag, a new one would be created.
A sparse collection can now contain Event Tags that use different
descriptors. \footnote{ Recall that this situation would force the current
system to use a single collection-owned descriptor, and thus bloat the
size of every tag, assuming the tag layouts were compatible in the first
place. }

\subsection{Eliminating Obsolete Data}
\subsubsection{General Approach}
As event store database developers, we do not have the
expertise to declare data obsolete, unneeded, or unnecessary.  However, we
can observe the system, and our observations point to a number of fields
and values which do not seem used.  BaBar physicists have confirmed that
many of these fields and values are, indeed, obsolete.  Thus the redesign
includes elimination of these fields.  Any field may be added or restored,
but the redesigned system will no longer include these unused fields by
default.

{\bf Tag Attributes} 
On average, more than 500 attributes are defined for each Event Tag in the
production system. We contacted the physics group who confirmed that several
attributes were obsolete. Also, some attributes were used during the initial
creation of events but were not needed for any subsequent filtering. In order
to handle such cases, the redesign will feature an updated Tag interface
that will allow the creation of \emph{transient-only} attributes that are
never stored in persistent Tags/Descriptors.

\section{Impact} \label{sec-impt}

Since 2000, BaBar disk space has been SLAC's largest single Computing
budget expense. The Event Store (navigational 
+ data components) accounts for
97\% of the 800TB+ BaBar database. Most of these files are migrated to tape
but Event Store navigation objects--the target of this redesign-- are the
most frequently accessed so they 
should be disk resident when possible.

Conservative estimates indicate that the redesign results in an 80\%
reduction in the navigation component size (2.2kB to 0.5kB per event).
Without benchmarks, performance gains are hard to quantify but using
fewer, smaller and less frequently accessed persistent objects will only
improve I/O latency. We look forward to running head-to-head comparisons.

\section{Project Status} \label{sec-status}
The implementation is progressing at a rapid pace. We will soon halt schema
development and focus on testing core functionality while establishing
backward compatibility guidelines to ensure that users can access data
produced by the original Event Store. Up to this point, we have kept our
code independent of the central BaBar development release cycle to avoid
disruption. The eventual merge will take place soon.

\section{Conclusion} \label{sec-concl}

The original BaBar Event Store was designed before its operation and use
were understood.  Its successful design has enabled it to well exceed the
original design requirements, scaling to a level beyond anyone's
expectations.  Using its generous flexibility, BaBar Event Store
developers were able to extend, modify and tune the system every step of
the way.  As the experiment continued and the Event Store grew,
accumulated experience showed room for improvement in size.  Now armed
with this experience, the redesign aims to dramatically reduce the Event
Store's persistent footprint.  

The Event Store Redesign project has succeeded in meeting its goals. The
overall size of the persistent event has been significantly reduced by
eliminating redundancy via common objects, removing obsolete data and
carefully re-organizing for more efficient access to persistent data. Most
importantly, backwards compatibility has been preserved.  

\begin{acknowledgments}
We wish to acknowledge David Quarrie and Simon Patton for their work on
the original BaBar event store~\cite{evsdesign-ref}.  With their
careful design, the system 
has been able to scale well beyond the original specifications.
\end{acknowledgments}


\end{document}